# Evolution of passive film behavior on mechanically polished NiTi shape memory alloy during forward martensitic transformation

Mahdi MOHAJERI[1], Homero CASTANEDA LOPEZ[2], Dimitris C LAGOUDAS[3]

[1] *Department of Materials Science & Engineering, Texas A&M University, College Station United States of America, mahdi_mohajeri@tamu.edu*

[2] *Department of Materials Science & Engineering, Texas A&M University, College Station United States of America, hcastaneda@tamu.edu*

[3] *Department of Aerospace Engineering, Texas A&M University, College Station, United States of America, lagoudas@tamu.edu*

Abstract:

The corrosion property of Nickel-titanium (NiTi) shape memory alloy is investigated during forward martensitic transformation between 40°C and 0°C. The Differential Scanning Calorimetry technique is used to find the forward and reverse martensitic transformation temperatures. NiTi shows the present of martensitic phase at near room temperature by cooling from 80°C. The change in corrosion behavior is monitored by electrochemical open circuit potential (OCP) during phase transformation. Cyclic potentiodynamic polarization method is used to identify the corrosion compound of NiTi. Corrosion rate at each temperature extracted from polarization data and activation energy for corrosion reaction in martensite and austenite phase are calculated based on an Arrhenius relation between corrosion current density and reciprocal temperature.

## Introduction

Nickel-Titanium (NiTi) shape memory alloys (SMAs) have attracted considerable interest for their special mechanical properties. Success of application of NiTi alloys in various biomedical and industrial areas is related to their functional properties (pseudo-elasticity and thermal shape memory). NiTi alloys are used in biomedical application for their good corrosion resistivity and biocompatibility. Many authors reported contradictory results for mechanically polished shape memory alloys. Wide variability of corrosion resistance from poor to excellent has been reported for dental orthodontic NiTi[1].
It is reported that heat treatment of NiTi implants with sterilization process can change the values of breakdown potential in compare with mechanically polished samples[2].

In this research, the forward phase transformation rule in corrosion property of NiTi is studied. Austenite phase transit to martensite phase during forward martensitic transformation with an exothermic reaction. The activation energy for both austenite and martensite phases can be calculated with an Arrhenius type equation between corrosion current density with reciprocal temperature.

## Experimental

### Materials
Nearly equiatomic NiTi alloy (50.8 at.%) was used for this study. All discs were mechanically polished through a series of silicon carbide papers up to a 2000-grit surface finish and then finished mirror-like using a 3-μm diamond paste.

### Differential Scanning Calorimetery (DSC)
A Perkin Elmer Pyris 1 Differential Scanning Calorimeter has been used to conduct the DSC analyses over a temperature range from -10°C to 200 °C. The liquid nitrogen cooling accessory is used to achieve subambient temperature. The linear heating or cooling rate is controlled at a standard rate of 10°C/min. Dry nitrogen at a rate of 20 ml/min has purged in the DSC cell. DSC apparatus has been calibrated with Indium.

### Potentiodynamic tests
Electrochemical experiments are performed in 3.5wt% NaCl solution on a computer-controlled potentiostat (Solartron 1285). Three electrode cell (Avesta Cell) is used for the tests with water circulator system to maintain temperature during heating and cooling. A graphite rod electrode and a saturated calomel electrode (SCE) are used as a counter electrode and as a reference electrode, respectively. The solution has been deaerated by purging nitrogen gas. All the potentials are reported with respect to SCE reference electrode. The open circuit potential has been recorded at test temperatures after 60 minutes that solution reaches each temperatures. Subsequently, the cyclic polarization scan is started at a 500 mV more cathodic potential than the OCP and increasing the anodic values with 0.17 mV/S up to 1700 mV/SCE.

## Results and Discussion

### Differential Scanning Calorimetry (DSC)
The transformation behavior of the microstructure during cooling and heating cycles has been measured using DSC method. Fig.1 shows DSC measurements of the transformation behavior starting from heating at room temperature. NiTi sample has been cooled in liquid nitrogen prior to eh DSC measurement. Two heating and cooling cycles has been done to find transformation temperatures. Table1 shows data of the DSC measurement.

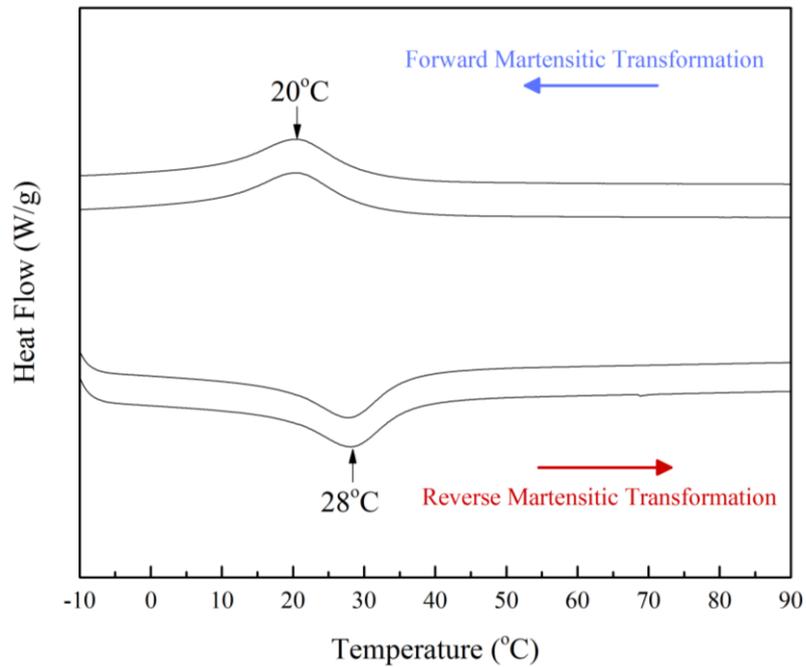

*Figure 1- DSC measurement of NiTi sample with 50.8 at.% Ni after polishing. Solid arrow lines show the thermal cycling directions for martensite transformation.*

Heat of reaction for the forward martensitic transformation is exothermic and for the reverse martensitic transformation is endothermic.

*Table 1- Transition temperatures of reversible forward and reverse martensitic transformation ($M_s$: Martensite start; $M_f$: Martensite finish; $M_p$: Martensite peak; $A_s$: Austenite start; $A_f$: Austenite finish; $A_p$: Austenite peak)*

| Reaction | $M_s$ (°C) | $M_f$ (°C) | $M_p$ (°C) | $A_s$ (°C) | $A_f$ (°C) | $A_p$ (°C) |
| --- | --- | --- | --- | --- | --- | --- |
| Forward transformation | 10 | 30 | 20 | ---- | ---- | ---- |
| Reverse Transformation | ---- | ---- | ---- | 22 | 38 | 28 |

The martensite start temperature ($M_s$) is near to room temperature that it means temperature near to room temperature during cooling cycle sample has mixture of martensite and austenite phases and during heating near to room temperature can find complete martensite phase.

Open Circuit Potential (OCP)

Figure 2 shows the evaluation on open circuit potential (OCP) during forward martensitic transformation. Prior to measurements, samples are heated up to 80°C in solution and then is cooled down to experiment temperature. With this preheat treatment in solution, the forward phase transformation follows the DSC trend and at 30 and 40°C NiTi sample has austenitic microstructure and at 0 and 10°C has martensitic microstructure. Martenstie peak at 20°C (Figure 1) indicate an exothermic reaction which it leads to highest corrosion potential in comparison with other temperatures.

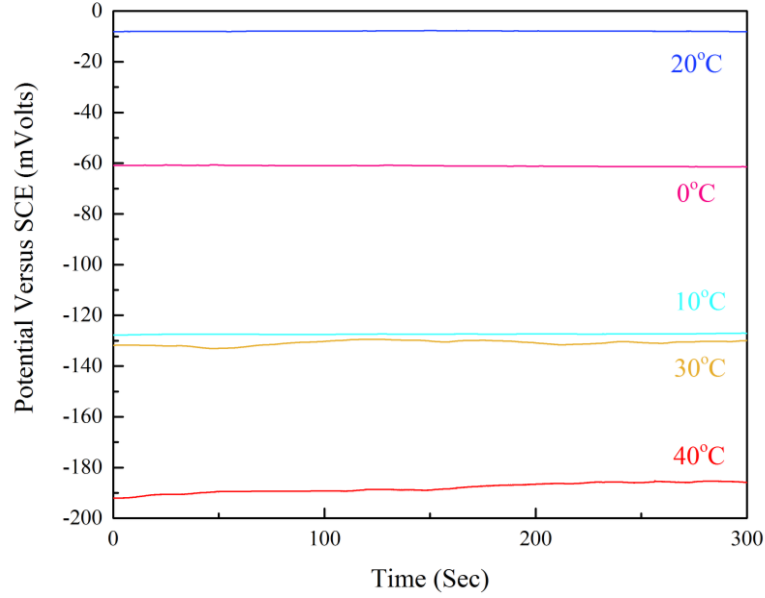

*Figure 2 - Variation of the OCP of the NiTi 50.8at.% in 3.5 wt% NaCl solution at 0, 10, 20 ,30 and 40 °C.*

Microstructure transforming from austenite to martensite lead to shifting the open circuit potentials to higher potentials.

Cyclic Potentiodynamic polarization

Fig.1 presents the cyclic polarization curves for NiTi specimens at various temperatures during forward martensitic phase transformation in 3.5wt% NaCl solution.

The cyclic polarization curve exhibits a negative hysteresis in cyclic polarization cures imply that pitting or crevice corrosion are not likely to occur at 0°C.

The polarization resistance ($R_P$) is used to measure corrosion rate near to corrosion potential ($E_{corr}$). $R_p$ is defined as the slope of the potential (E) versus current density (i) plot near the open circuit potential by Eq 1 with the potential range 30mV on cathodic and anodic side of the open circuit potential [3, 4].

$$R_P = (\frac{\partial \Delta E}{\partial i})_{at\ t=0,\ \Delta E=0} \ \Omega.cm^{-2} \qquad 1$$

And corrosion current density ($i_{corr}$) is described by Stern-Geary expression (Eq 2).

$$i_{corr} = \frac{1}{2.3R_P}(\frac{\beta_a \beta_c}{\beta_a + \beta_c}) \qquad 2$$

Where $\beta_a$ and $\beta_c$ are the anodic and cathodic Tafel slopes, respectively.

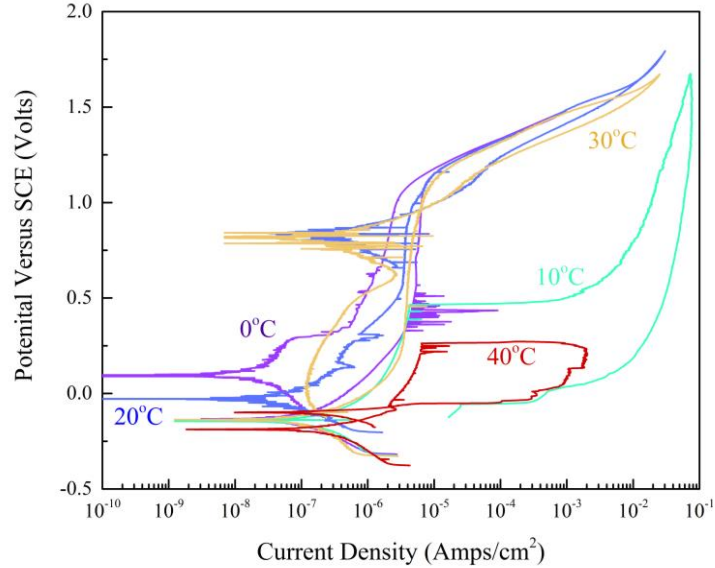

*Figure 3- Representative Cyclic polarization curves of NiTi exposed to 3.5wt% NaCl solution at different temperatures (During forward martensitic transformation)*

Tafel slopes are determined from the polarization curves in potential near to open circuit potential with low voltage scan rate (to obtain near steady state condition) by Eq 3[5].

$$i_{app} = i_{corr}(\exp[\frac{2.3(E-E_{corr})}{\beta_a}] - \exp[\frac{2.3(E-E_{corr})}{\beta_c}]) + c(\frac{\partial E}{\partial t}) \qquad 3$$

Where E and $E_{corr}$ are applied potential and corrosion potential, respectively. "c" is the interfacial capacitance related to the electrochemical double layer. The second term in right hand side of equation approaches zero in low scan rate.

The non –linear least square Levenburg-Marquardt (LEV) fit method is used to perform the Tafel slopes values[6].

The corrosion parameters derived from the cyclic potentiodynamic polarization data are summarized in Table I.

*Table 2 – Results of cyclic potentiodynamic polarization measurement of NiTi alloy in 3.5 wt% NaCl solutions at different temperatures.*

| Temperature | Phase | $E_{corr}$ (mV/SCE) | $I_{corr}$ ($\mu A\ cm^{-2}$) | $\beta_a$ (mV/dec) | $\beta_c$ (mV/dec) | $R_P$ ($k\Omega cm^2$) |
|---|---|---|---|---|---|---|
| 40 | A | -187.9 | 0.51 | 126.4 | 279.5 | 860 |
| 30 | A | -137.7 | 0.33 | 260.5 | 574.1 | 219 |
| 20 | A+M | -28.44 | 0.08 | 208.2 | 242.1 | 648 |
| 10 | M | -146.4 | 0.81 | 266.3 | 364.7 | 162 |
| 0 | M | -136.3 | 0.17 | 279.5 | 234.6 | 327 |

M : Martensite ; A: Austenite

In general, the change in corrosion rate which is represented by corrosion current density ($i_{corr}$) with temperature follow Arrhenius Eq 4[7-9]:

$$\log i_{corr} = \log A - \frac{E_a}{2.303RT} \qquad 4$$

Where $E_a$ (J/mol) is the molar activation energy of the process, $R$ is the gas constant (8.314 J/mol/K) and $A$ is a constant. The relation between log $i_{corr}$ and reciprocal temperature (1/T) is presented in Figure 4.

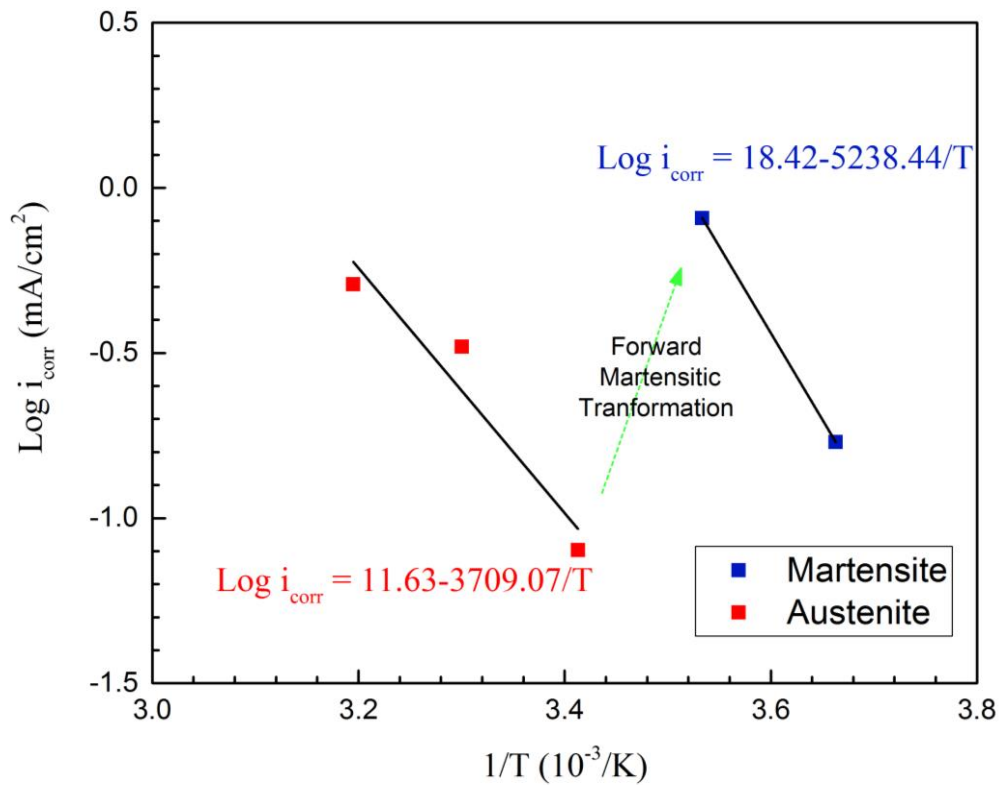

*Figure 4- Corrosion current density vs. time during forward martensitic phase transformation.*

The activation energy refers to the energy level that must be overcome by electron through the electrode and electrolyte interface. Linear relation between logarithmic current density and reciprocal temperature is represent for martensite and austenite microstructure and clearly the change in corrosion mechanism can be found with change in activation energy of corrosion process between the austenite and martensite phases.

Conclusion

This study has shown a change in corrosion resistance of NiTi alloy during forward martensitic transformation from austenitic phase to martensitic phase after exposure in 3.5wt% NaCl solution. The austenite phase has lower activation energy in compare to martensite phase which it means electron in metal/electrolyte interface need to overcome lower energy level. In temperature related to peak of martensite transformation lowest corrosion rate and higher open circuit potential indicate the change in corrosion site and mechanism on the surface of metal.